\documentclass[conference]{IEEEtran}
\usepackage{graphics}                 
\usepackage{graphicx}
\usepackage[cmex10]{amsmath}
\usepackage{amssymb, amsfonts}
\usepackage{algorithmic}
\usepackage{algorithm}
\usepackage{cite}
\usepackage{verbatim}
\usepackage{url}
\ifCLASSOPTIONcompsoc
\usepackage[tight,normalsize,sf,SF]{subfigure}
\else
\usepackage[tight,footnotesize]{subfigure}
\fi
\usepackage{subfigure}

\graphicspath{{figures/}}

\begin{document}
%
\title{A Joint Model for IEEE 802.15.4 Physical and Medium Access Control Layers}

\author{\IEEEauthorblockN{Mohamed-Haykel Zayani\IEEEauthorrefmark{1}, Vincent Gauthier\IEEEauthorrefmark{1}, Djamal Zeghlache\IEEEauthorrefmark{1}}
\IEEEauthorblockA{\IEEEauthorrefmark{1}Lab. CNRS SAMOVAR UMR 5157,
Telecom SudParis, Evry, France} \IEEEauthorblockA{Emails:
\{mohamed-haykel.zayani, vincent.gauthier,
djamal.zeghlache\}@telecom-sudparis.eu}
 }


%


\IEEEoverridecommandlockouts


\IEEEpubid{\makebox[\columnwidth]{\hfill
978-1-4244-9538-2/11/\$26.00~\copyright~2011 IEEE} \hspace
{\columnsep}\makebox[\columnwidth]{ }}

\maketitle

\begin{abstract}
Many studies have tried to evaluate wireless networks and especially
the IEEE 802.15.4 standard. Hence, several papers have aimed to
describe the functionalities of the physical (PHY) and medium access
control (MAC) layers. They have highlighted some characteristics
with experimental results and/or have attempted to reproduce them
using theoretical models. In this paper, we use the first way to
better understand IEEE 802.15.4 standard. Indeed, we provide a
comprehensive model, able more faithfully to mimic the
functionalities of this standard at the PHY and MAC layers. We
propose a combination of two relevant models for the two layers. The
PHY layer behavior is reproduced by a mathematical framework, which
is based on radio and channel models, in order to quantify link
reliability. On the other hand, the MAC layer is mimed by an
enhanced Markov chain. The results show the pertinence of our
approach compared to the model based on a Markov chain for IEEE
802.15.4 MAC layer. This contribution allows us fully and more
precisely to estimate the network performance with different network
sizes, as well as different metrics such as node reliability and
delay. Our contribution enables us to catch possible failures at
both layers.
\end{abstract}

\IEEEpeerreviewmaketitle


\section{Introduction}
Wireless sensor networks have been closely studied in recent years.
Several studies have investigated behaviors and performances of
these networks. Some of them have highlighted such networks
properties by relying on empiric results
\cite{Cerpa2003,Zhao2003,Woo2003,Ganesan,Cerpa2005} whereas others
have focused on reproducing a standard or mechanism functionalities
tied to sensors by proposing analytical models
\cite{Bianchi2000,Zhai2004,Daneshgaran2008a,Park2009}. Empirical
studies have shown that wireless communication networks are
radically different from some simulation models (disc-shaped nodes
range for example). Analytical studies have attempted to reproduce
mechanisms and technical aspects widely used/seen in these networks
in order to track network performances. Among these approaches,
those of Zuniga and Krishchnamachari \cite{Zuniga2004,Zamalloa2007}
stand out. They emphasize the limits of disc-shaped node range
models that are used in simulators, and highlight the existence of a
transitional region between the connected and disconnected areas.
This observation, based on experiments, enables us to understand
clearly the reason behind link unreliability in low power wireless
networks. Moreover, Zuniga and Krishchnamachari underline the impact
of asymmetry in transitional region expansion and its negative
effect on reliability \cite{Zamalloa2007}. Meanwhile, lot of the
performance analysis of MAC layer protocol are derived from the
Markov model proposed by Bianchi \cite{Bianchi2000} for the IEEE
802.11 standard \cite{IEEE80211}. This model consists in a Markov
chain that reproduces the functionalities of the IEEE 802.11
standard while assuming saturated traffic and ideal channel
conditions. This approach has inspired many others, for instance the
Park et al. model \cite{Park2009}. It presents itself as a relevant
contribution which aims to measure reliability, delay and energy
consumption in a wireless network based on IEEE 802.15.4 standard
\cite{IEEE802154}. In this paper, we develop an IEEE 802.15.4 model
that takes into consideration the interactions on PHY and MAC
layers. The model, at the PHY layer level, is derived from the
Zuniga and Krishnamachari mathematical framework for quantifying
link unreliability. The MAC layer model is inspired from an enhanced
Park et al. Markov chain. The joint model that combines both PHY and
MAC models enables us to consider the causes behind packet losses
either at PHY or MAC levels. Indeed, in the Park et al. model,
collisions appear to be the only reason for losses, whereas, our
model includes constraints posed by SNR (signal-to-noise ratio),
modulation, encoding and asymmetry (heterogeneous hardware).

The remainder of this paper is as follows. In Section II, we present
the related work which gives an overview of approaches that inspire
our model. We focus on our contribution by giving details on the
combined PHY and MAC layers models in Section III. In Section IV, we
compare our proposition to the enhanced Park et al. approach and
estimate nodes performances with different network densities.
Finally, Section V concludes the paper and discusses some future
research challenges. \IEEEpubidadjcol

\section{Related Work}
Many studies have aimed to understand and to evaluate standards and
protocols. The works that tried to identify the properties of these
networks mechanisms fall into two categories: i.e. simulations-based
\cite{Cerpa2003,Zhao2003,Woo2003,Ganesan,Cerpa2005} relying on
empiric observations, or analytical works
\cite{Bianchi2000,Zhai2004,Daneshgaran2008a,Park2009}. Most of
analytical studies are based on the Markov model proposed by Bianchi
\cite{Bianchi2000} for the IEEE 802.11 standard. This model consists
in a Markov chain that mimics the functionalities of the IEEE 802.11
standard while assuming a saturated traffic and ideal channel
conditions. Zhai et al. \cite{Zhai2006a} and Daneshgaran et al.
\cite{Daneshgaran2008a} exploit the Bianchi model and extend it
through more realistic assumptions. These approaches have inspired
Griffith and Souryal \cite{NIST} to develop a model for the IEEE
802.11 MAC layer that adds a frame queue to each node. This
contribution enables us to estimate the packet reception rate, the
delay, the medium access control layer (MAC layer) service time and
the throughput. Similar studies have been developed for the wireless
sensor networks, and more especially the IEEE 802.15.4 standard.
Hence, we note the models developed by Pollin et al.
\cite{Pollin2008} and Park et al. \cite{Park2009}. The two
approaches provide a generalized analysis that allows to measure
reliability, delay and energy consumption. In each proposed model,
the exponential backoff process is modeled by a Markov chain. Retry
limits and acknowledgements in an unsaturated traffic scenario are
also taken into consideration.

Park et al. propose a generalized analytical model of the slotted
CSMA/CA mechanism with beacon enabled mode in IEEE 802.15.4. This
model includes retry limits for each packet transmission. The
scenario of a star network in which $N$ nodes try to send data to a
sink has been considered and defining the state of a single node
through a Markov model has been proposed. Each state of the Markov
chain is characterized by three stochastic processes: the backoff
stage $s(t)$, the state of the backoff counter $c(t)$ and the state
of the retransmission counter $r(t)$ at time $t$. The described
modeling allows us to analyze of the link reliability, delay and
energy consumption.

In another context, numerous works focus on the physical layer
modeling. For instance, Zuniga and Krishnamachari
\cite{Zuniga2004,Zamalloa2007} have analyzed the major causes behind
unreliability \cite{Zuniga2004,Zamalloa2007} and the negative impact
of asymmetry on link efficiency \cite{Zamalloa2007} in low power
wireless links. Instead of the binary disc-shaped model these models
reproduce the called $transitional$ $region$
\cite{Zhao2003,Woo2003,Ganesan} in order to model the transmission
range. The packet reception rate and the upper-layer protocol
reliability are highly instable when a neighbor is located in this
region. To understand it, two models have been proposed: a channel
model that is based on the log-normal path loss propagation model
\cite{Rappaport2002} and a radio reception model closely tied to the
determination of packet reception ratio. Through these models, it is
possible to derive the expected distribution and the variance of the
packet reception ratio according to the distance.

\section{Developed IEEE 802.15.4 Model for Smart Grid}
Our contribution joins the initiative of Smart Grid \cite{NIST} to
provide tools that evaluate wireless communications standards. The
developed model that we propose analyzes an IEEE 802.15.4 PHY and
MAC layer channel in which multiple non-saturated stations compete
in communicating with a sink. The aim is to combine two relevant
models: A PHY model that bypasses the disk shaped node range and
takes into consideration the called transitional region and a MAC
model that reproduces the CSMA/CA mechanism. The model described
enables us to add the impact of PHY layer errors onto the MAC layer
and to provide some improvements for the adopted MAC model, in order
to obtain more precise output estimations. Our developed model is
available at the SGIP NIST Smart Grid Collaboration website
\cite{NIST}.

\subsection{IEEE 802.15.4 PHY Model Description}
To model the PHY layer, we have adopted the Zuniga and
Krishnamachari approach \cite{Zuniga2004,Zamalloa2007}. The main
objective is to identify the causes of the transitional region and
quantify their influence on performance without considering
interferences (assumption of a light traffic or static
interference). To do this, the expressions of the packet reception
rate as function of distance are derived. These expressions take
into account radio and channel parameters such as the path loss
exponent (log-normal shadowing path loss model
\cite{Rappaport2002}), the channel shadowing variance, the
modulation, the coding and hardware heterogeneity. They describe how
the channel and radio influence transitional region growth. We use
mathematical frameworks provided to compute packet delivery rate
independently of interferences. For more details of the Zuniga and
Krishnamachari models see \cite{Zuniga2004,Zamalloa2007}.

We believe that including the PHY model in the MAC model considered
(the next subsection describes the MAC model) is an interesting
challenge. Indeed, collisions are the major factor behind frame
losses. Nonetheless, considering errors that can happen at the PHY
layer includes constraints imposed by SNR (signal-to-noise ratio),
modulation, encoding and asymmetry (heterogeneous hardware).
Therefore, our contribution allows us to have a more realistic
estimation by taking into account the causes of failures at both
layers.

\subsection{Operation details for the IEEE 802.15.4 MAC Model and the interactions with the PHY Model}
The model of IEEE 802.15.4 MAC layer is inspired from Park et al.
Markov chain \cite{Park2009}. It captures the state of the station
backoff stage, the backoff counter and the retransmission counter.
We insert into Park et al. model an M/M/1/K queuing model that
endows a finite buffer to a station. On the one hand, the Markov
model determines the steady state probability when a station senses
the channel in order to transmit a frame and the probability that a
frame experiences a failure (due to a collision or to PHY layer
failure). On the other hand, the queuing model gives as output some
measurements such as the throughput or the probability that the
station is idle.

\begin{figure}
\begin{center}
\includegraphics[width=0.5\textwidth]{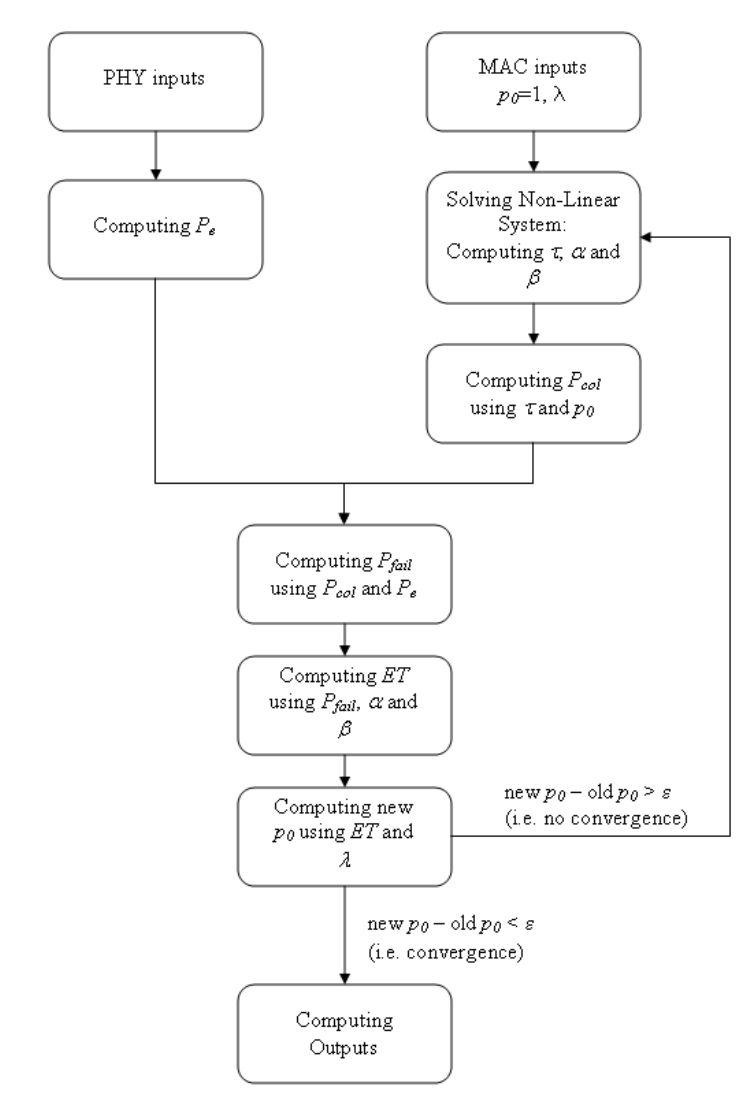}
\caption{IEEE 802.15.4 PHY and MAC model flowchart}
\label{Flowchart}
\end{center}
\end{figure}

The Park et al. approach, inspired from \cite{Pollin2008}, consists
in a generalized analytical model of the slotted CSMA/CA mechanism
of beacon enabled IEEE 802.15.4 with retry limits for each packet
transmission (the complete description is provided in
\cite{Park2009}). The model takes the scenario of $N$ stations that
try to communicate with a sink. Park et al. define the probabilities
for the following events: a node attempts a first carrier sensing to
transmit a frame, a node finds the channel busy during CCA1 or a
node finds the channel busy during CCA2. They are denoted by the
variables $\tau$, $\alpha$ and $\beta$ respectively. These three
probabilities are related by a system of three non-linear equations
that arises from finding the steady state probabilities. Our model,
described by the flowchart presented in Fig. \ref{Flowchart} (the
main PHY and MAC inputs are listed in Table I and Table II
respectively), aims to solve the non-linear system that expresses
these probabilities. In addition, it estimates $p_0$, the
probability of going back to the idle state by considering the
offered load per node parameter $\lambda$. In this way, our
contribution enables the MAC model to determine this probability, in
opposition to \cite{Park2009} ($p_0$ is taken as an input for the
performances analysis).

We start from equations (16), (17) and (18) in \cite{Park2009} and
make changes in some of these expressions to enhance the model. The
equations (17) and (18) are expressed with probability $\tau$ to
mention that a node is transmitting. In our mind, this consideration
is insufficient because a transmitting node must not be idle, that
is why we substitute $\tau$ by (1-$p_0$)$\tau$. Thereby, $\tau$ is
the probability that a node tries to transmit and 1-$p_0$ is the
probability that a station has a frame to send. The system
considered is given by equations (\ref{formula1}), (\ref{formula2})
and (\ref{formula3}):

\begin{eqnarray}
    \tau & = & \left(\frac{1-x^{m+1}}{1-x}\right)\left(\frac{1-y^{n+1}}{1-y}\right)b_{0,0,0} \label{formula1}\\
    \alpha & = & \left( L + \frac{N(1-p_0)\tau(1-\tau(1-p_0))^{N-1}}{1-(1-\tau(1-p_0))^N}L_{ACK}  \right) \nonumber\\
    & & \left( 1-(1-(1-p_0)\tau)^{N-1}  \right) (1-\alpha)(1-\beta) \label{formula2}\\
    \beta  & = & \frac{1-(1-\tau(1-p_0))^{N-1}}{DV} +  \nonumber\\
    & & \frac{N (1-p_0)\tau(1-(1-p_0)\tau)^{N-1}} {DV} \label{formula3}
\end{eqnarray}
Where $DV=2-(1-(1-p_0)\tau)^N+\\N(1-p_0)\tau(1-(1-p_0)\tau)^{N-1}$,
$x=\alpha+(1-\alpha)\beta$ and $y=P_{fail}(1-x^{m+1})$. The
parameter $P_{fail}$ represents the probability of a failed
transmission attempt, $m$ is the maximum number of backoffs the
CSMA/CA algorithm will attempt before declaring a channel access
failure, $n$ is the maximum number of retries allowed after a
transmission failure, $L$ is the length of the data frame in slots
(a slot has a length of 80 bits), $L_{ACK}$ is the length of an
acknowledgement in slots, $N$ is the number of stations and
$b_{0,0,0}$ is the state where the state variables of the backoff
stage counter, the backoff counter and the retransmission counter
are equal to 0 (an approximation is proposed in \cite{Park2009}).

The mechanism that computes these probabilities (using the MATLAB
$fsolve$ function) allows us to determine the probability of failed
transmission $P_{fail}$, given by:

\begin{equation}
    P_{fail} = 1-(1-P_{col})(1-P_e)
    \label{form3}
\end{equation}
Where $P_{col} = 1-(1-\tau(1-p_0))^{N-1}$.

In the above expressions, $P_e$ is the probability of loss due to
channel and radio constraints (computed by the PHY model) and
$P_{col}$ is the probability of a collision occurring (modified as
done with (17) and (18) in \cite{Park2009}).

This mechanism is embedded in a loop that updates $p_0$. The
developed model solves the system of non-linear equations to
determine $\tau$, $\alpha$, $\beta$ and therefore $P_{fail}$. Then,
$P_{fail}$, $\alpha$ and $\beta$ are used to estimate the mean MAC
service time, or the mean time to process a frame, expressed also as
Expected Time or $ET$ (in \cite{Park2009}, Section V.B details how
to compute this time. We substitute, of course, $P_{col}$ by
$P_{fail}$ to catch errors that can occur at PHY and MAC layers).
So, a new value for $p_0$ is generated and the updated $p_0$ is used
in the next iteration. It is possible to determine $p_0$ since each
device has a buffering capacity. Every node is modeled as an M/M/1/K
queue and each queue receives frames following a Poisson arrival
process $\lambda$ frames/s. The queue utilization $\rho$ is the
product of the arrival rate $\lambda$ and the inverse of the mean
MAC service time $ET$. The steady state probability that there are
$i$ frames in the queue is:

\begin{equation}
    p_i=\rho ^{i}/\sum_{j=0}^{K}\rho ^{j}
    \label{form4}
\end{equation}

Hence, the value of $p_0$ is given by:

\begin{equation}
    p_0=\left [ \sum_{j=0}^{K}\rho ^{j} \right ]^{-1}
    \label{form5}
\end{equation}

The process continues until the value of $p_0$ converges to a stable
value. Once $p_0$ converges, all outputs concerning queuing analysis
can be computed for each value of $\lambda$ (the per-node load
offered). Four outputs are considered in this study: the average
waiting time to receive a frame (Eq. (7)), the failure probability
(probability of packet loss due to collisions or link
constraints)(Eq. (4)), the reliability of a node (the probability of
a good frame reception)(Eq. (8)) and the average throughput per
node(Eq. (9)).

\begin{equation}
    D=\frac{L}{\lambda \left ( 1-p_{k} \right )}
    \label{form6}
\end{equation}
\begin{equation}
    R=\left (1-p_k  \right )\left (1-P_{cf}  \right )\left (1-P_{cr}  \right )
    \label{form7}
\end{equation}
\begin{equation}
    S_{avg}=\lambda RL_{p}
    \label{form8}
\end{equation}

Where $p_k$ is the probability of having full buffer, $P_{cf}$ is
the probability that the frame is discarded due to channel access
failure (Eq. (19) in \cite{Park2009}), $P_{cr}$ is the probability
that the packet is discarded due to retry limits (Eq. (20) in
\cite{Park2009}), $L$ is the payload size and $L_p$ is the
application data size.

Therefore, this contribution enables us to enhance Park and al.
model at two levels:
\begin{itemize}
\item Providing a more precise computation of failure probability by considering possible errors at PHY and MAC layers (link unreliability and collisions).
\item Enhancing the MAC model of Park et al. by estimating the probability $p_0$ for the resolution of non-linear equations (this probability is an input in Park et al. model), modifying some expressions to more efficient estimations and determining outputs relative to a precise scenario (number of nodes and per-node load offered).
\end{itemize}

\section{Simulation and Results}
We propose two scenarios for the simulations. Firstly, we compare
the performances of a node obtained in two different ways. On one
hand, we use the Park et al. Markov chain (MAC layer) and on the
other hand we test our model. Secondly, we expose the same
performances, using our developed model, for different densities.
Table \ref{table1} presents the main inputs at the MAC layer and
Table \ref{table2} enumerates the main ones at the PHY layer. All
the simulations test different values for the offered per-station
load, measured in units of frame/s. We choose to start from 0.5
frame/s and increase the offered load to 25 frames/s with a step of
0.5 frame/s (or from 400 bits/s to 20000 bits/s). We select four
outputs to illustrate node performances: the average waiting time
for a frame reception, the failure probability (probability of frame
loss due to collisions or link constraints), the reliability of a
node (the probability of a good frame reception) and the throughput.

\begin{table}[!t]
\renewcommand{\arraystretch}{1.3}
\caption{Main Parameters used in MAC Layer}
\label{table1}
\centering
\begin{tabular}{c||c||c||c}
\hline
\bfseries Parameter & \bfseries Value & \bfseries Parameter & \bfseries Value\\
\hline\hline
Number of Nodes & 5, 10, 50 & Queue Size & 51 Frames \\
Smallest Backoff Win & 8 & Data Rate & 19.2 kbit/s \\
Max Frame Retries & 3 &  ACK Size & 88 bits \\
Max CSMA Backoff & 4 & Shadowing STD & 4 \\
Max Backoff Exponent & 5 & IFS & 640 $\mu$s \\
Min Backoff Exponent & 3 & Max TX-RX Time & 192 $\mu$s \\
MAC Frame Payload & 800 bits & MAC Overhead & 48 bits\\
\hline
\end{tabular}

\end{table}
\begin{table}[!t]
\renewcommand{\arraystretch}{1.3}
\caption{Main Parameters used in PHY Layer}
\label{table2}
\centering
\begin{tabular}{c||c||c||c}
\hline
\bfseries Parameter & \bfseries Value & \bfseries Parameter & \bfseries Value\\
\hline\hline
Noise Figure & 23dB & Bandwidth & 30kHz \\
Pathloss exp & 4 & STD Tx power & 5dBm \\
Noise & 15dB &  Preamb. Length & 40 bits \\
Max Tx range & 20 m & Min Tx Range & 1 m \\
\hline
\end{tabular}
\end{table}

\begin{figure*}[!tb]
  \centering
  \subfigure[Average wait time versus load offered]{
    \label{fig1}
    \includegraphics[width=0.45\textwidth]{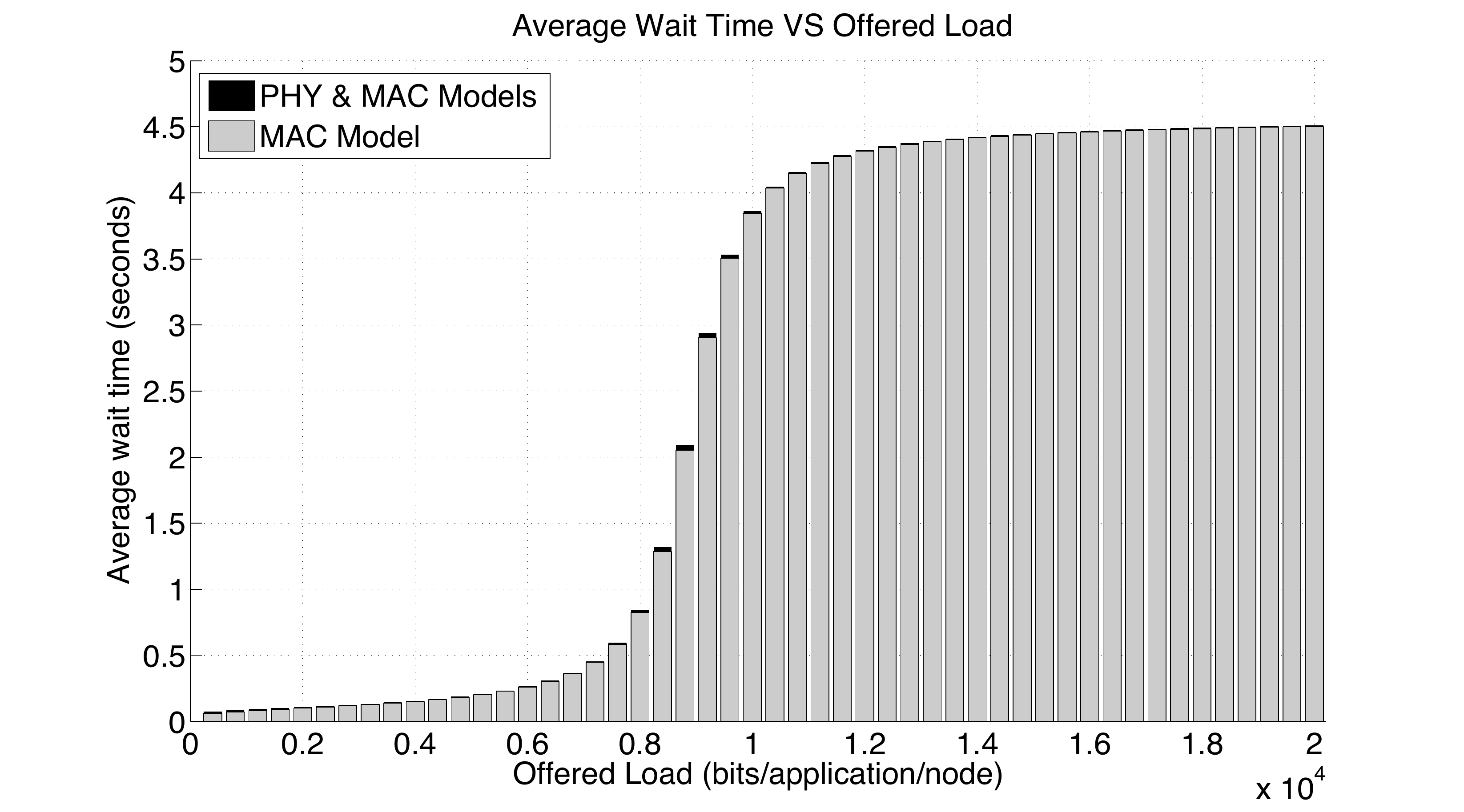}
  }\hfill
  \subfigure[Failure probability versus load offered]{
    \label{fig2}
    \includegraphics[width=0.45\textwidth]{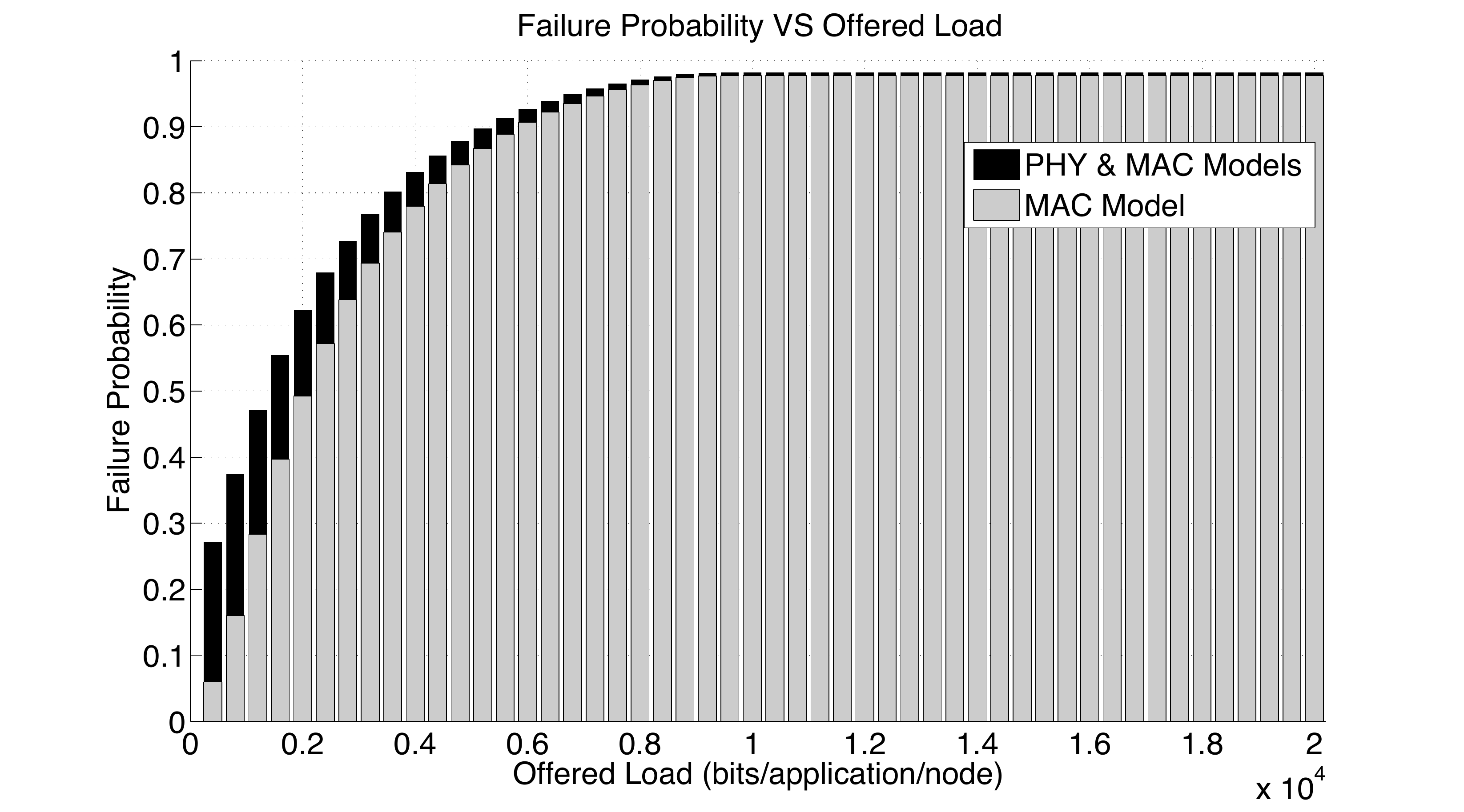}
  }\\
 \subfigure[Reliability versus load offered]{
    \label{fig3}
    \includegraphics[width=0.45\textwidth]{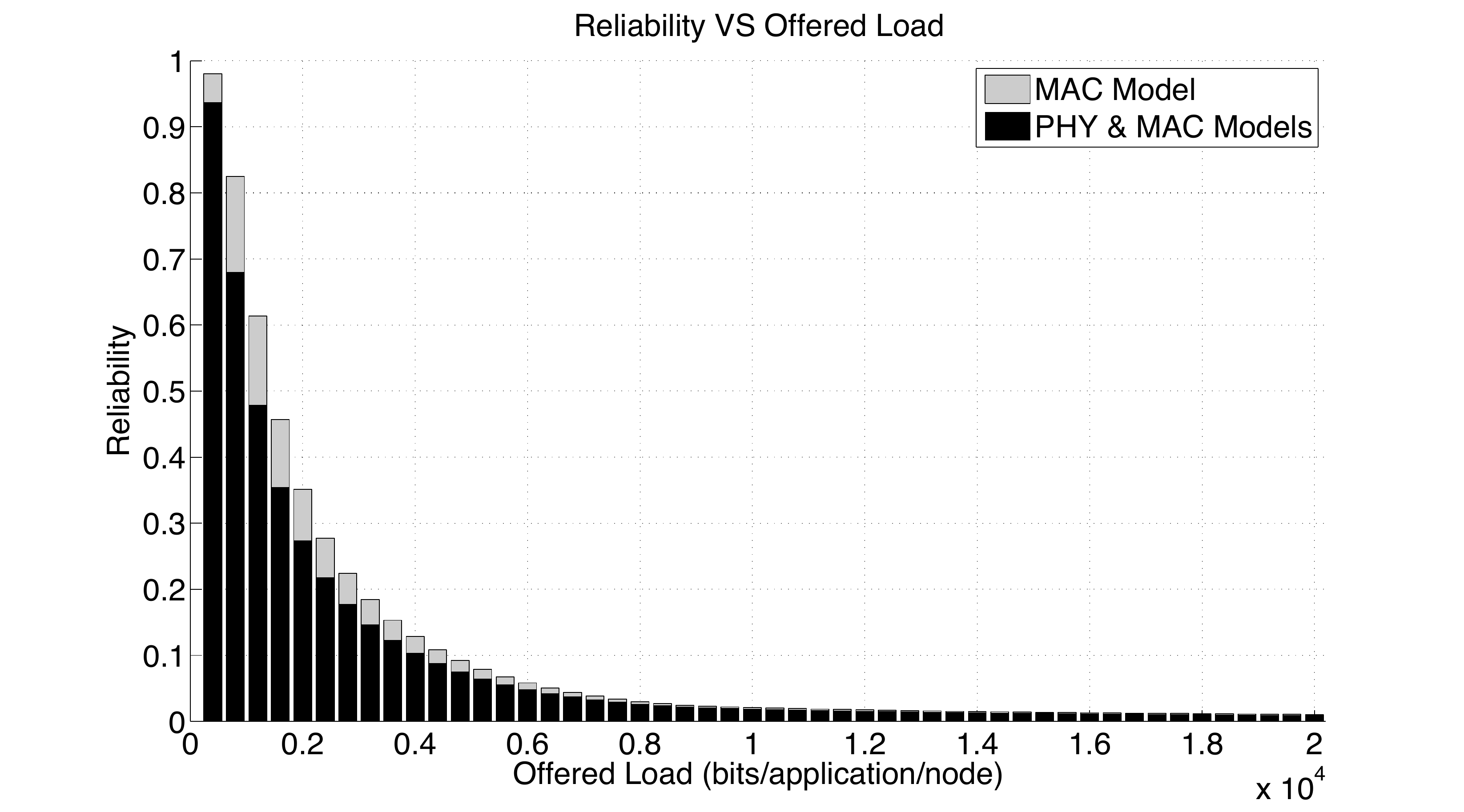}
  }\hfill
  \subfigure[Throughput versus load offered]{
     \label{fig4}
     \includegraphics[width=0.45\textwidth]{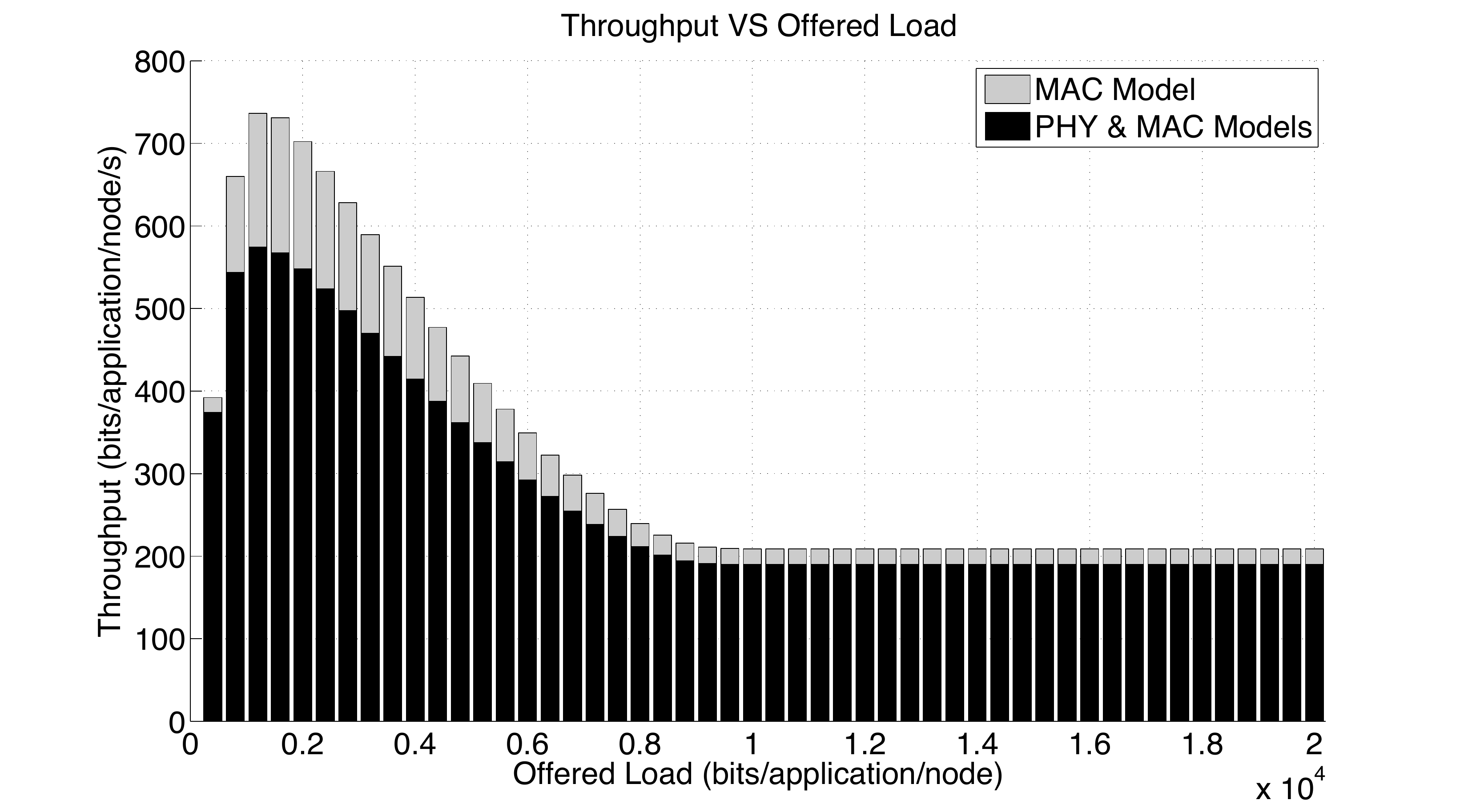}
  }
  \caption{Comparison between IEEE 802.15.4 PHY \& MAC Model and IEEE 802.15.4 MAC Model}
  \label{MACPHY}
\end{figure*}

\begin{figure*}[!tb]
  \centering
  \subfigure[Average wait time versus load offered]{
    \label{fig5}
    \includegraphics[width=0.45\textwidth]{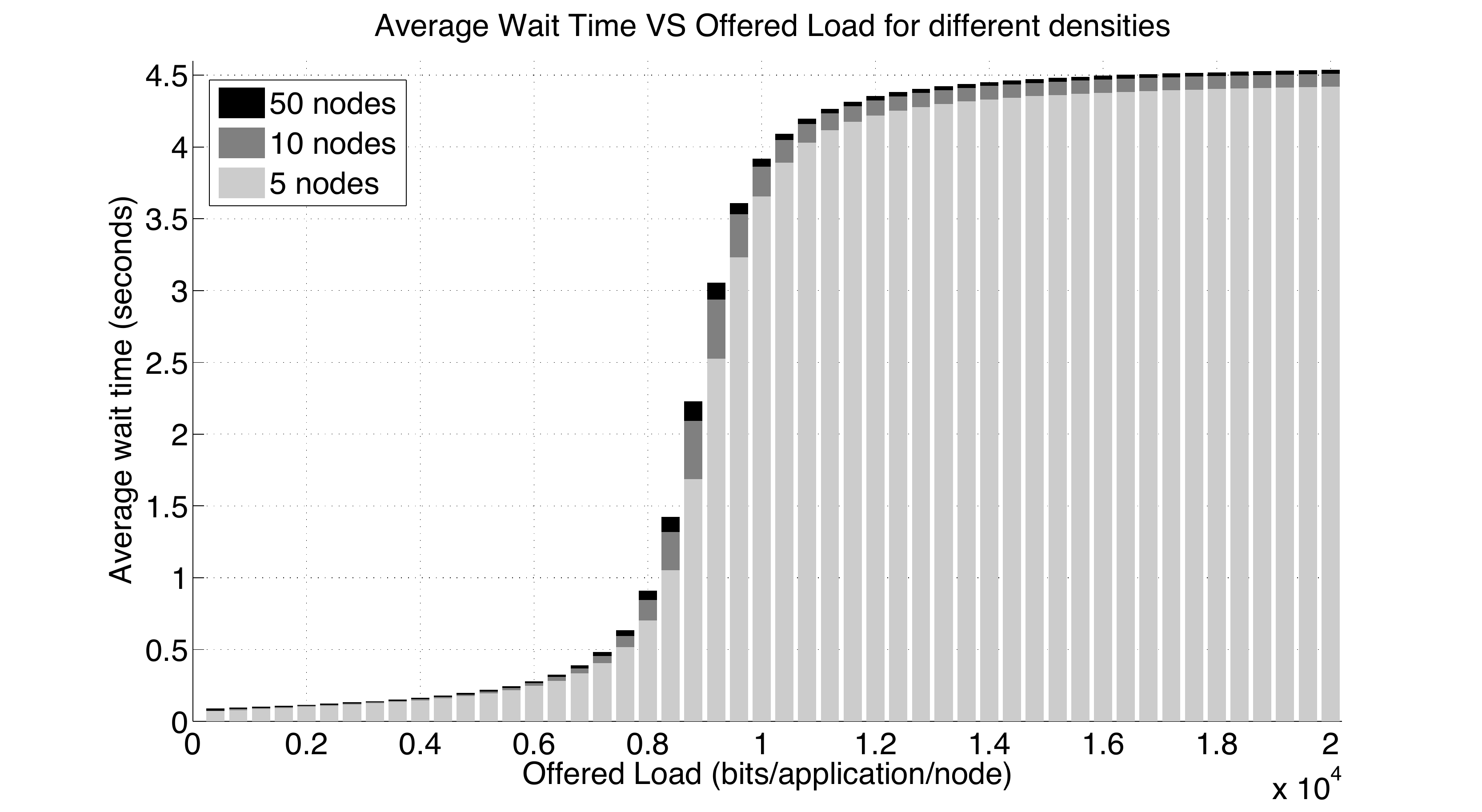}
  }\hfill
  \subfigure[Failure probability versus load offered]{
    \label{fig6}
    \includegraphics[width=0.45\textwidth]{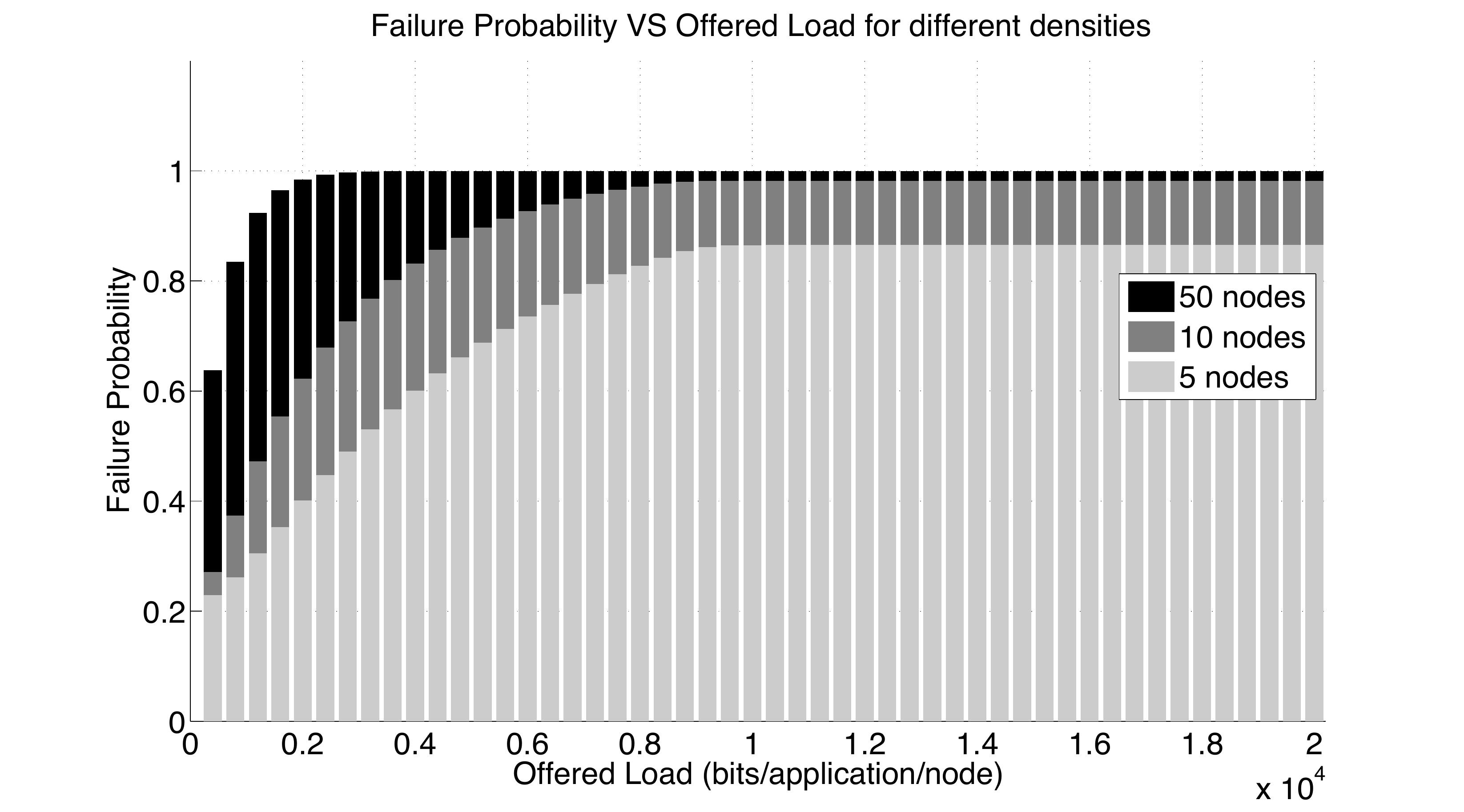}
  }\\
 \subfigure[Reliability versus load offered]{
    \label{fig7}
    \includegraphics[width=0.45\textwidth]{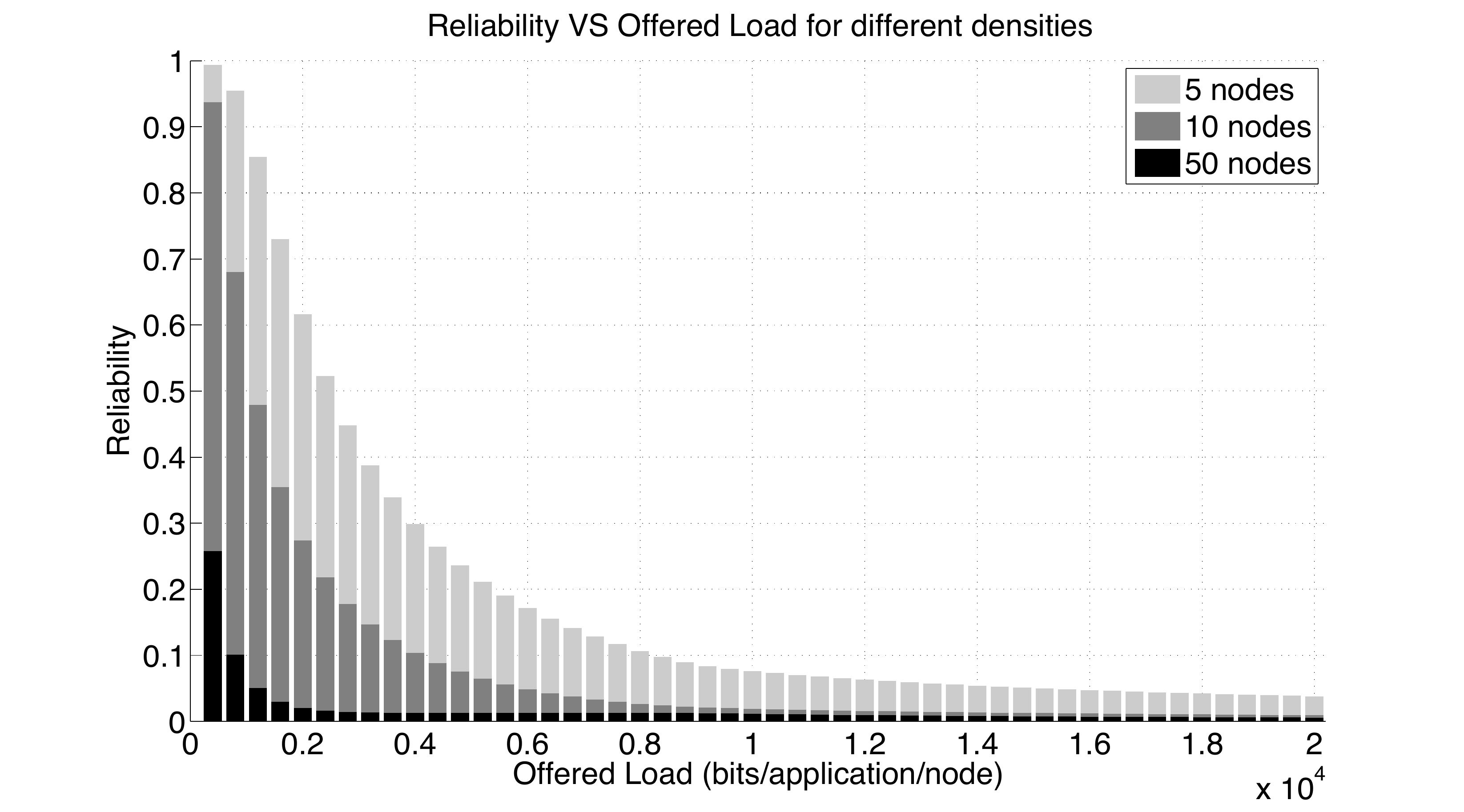}
  }\hfill
  \subfigure[Throughput versus load offered]{
     \label{fig8}
     \includegraphics[width=0.45\textwidth]{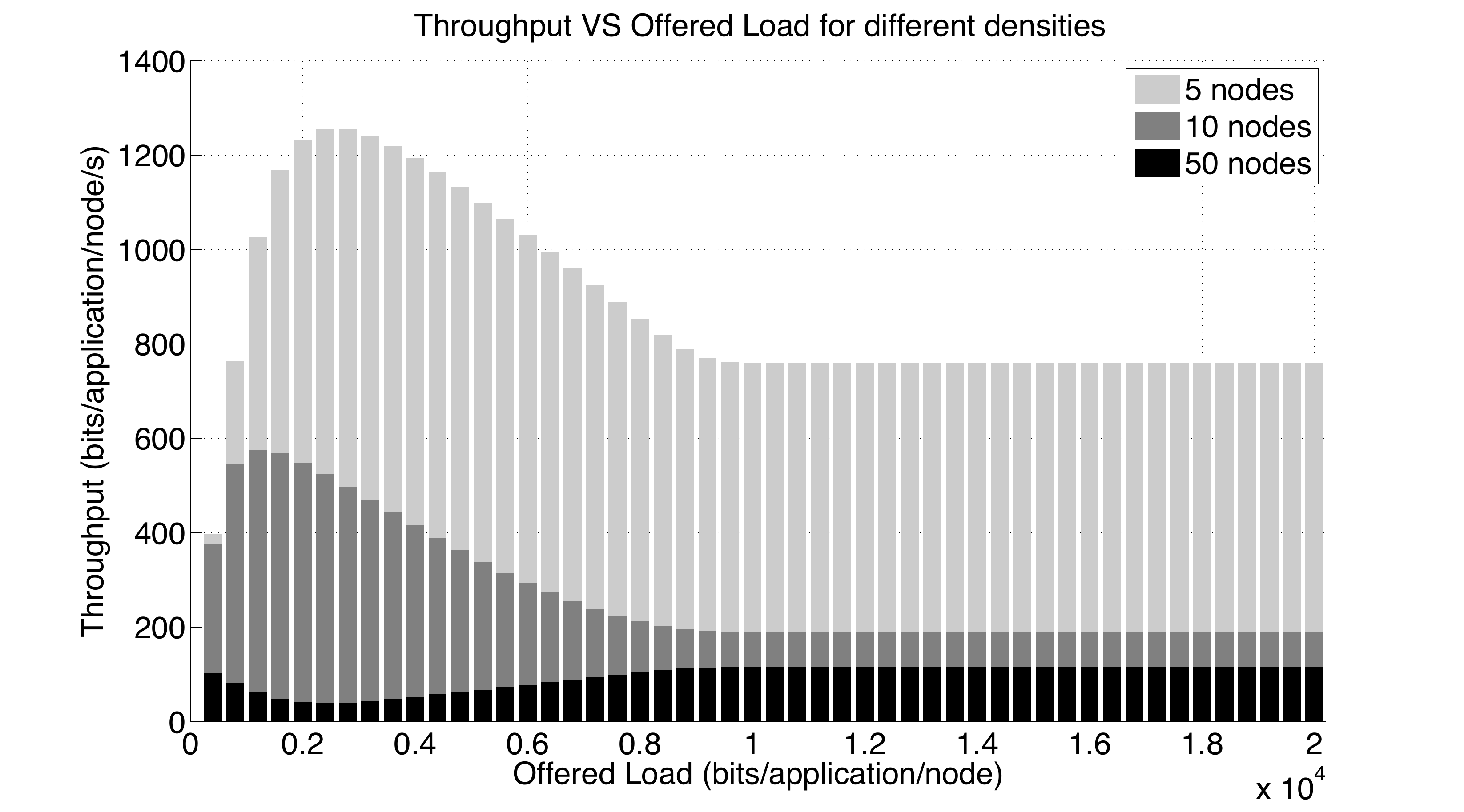}
  }
  \caption{Performances evolutions for different densities using IEEE 802.15.4 PHY \& MAC Model}
  \label{perf}
\end{figure*}

\subsection{Comparison between combined PHY and MAC layers and simple MAC layer models}
As described in Section III, when we include the constraints at the
physical layer, delivery failures happen more often. There are many
reasons for this: weak SNR and modulation and/or encoding errors. We
run simulation for a star network with 10 nodes. The results confirm
a notable degradation of node performances. In Fig. \ref{fig1}, the
average waiting time is for the the combination of the PHY and MAC
models. Inserting link constraints increases the number of
retransmissions. Thus, the delay increases. The delay difference
between the two approaches reaches 40 ms at offered load equal to 11
frames/s. Fig. \ref{fig2} compares the evolution of failure
probability for the two approaches. With light offered loads, the
impact of the PHY model is conspicuous, especially since the number
of collisions is likely to be low. The collisions are more frequent
with heavier loads and the probability of occurrences grows quickly,
generating network saturation. Meanwhile, the probability of losses
due to link conditions remains constant (this probability is
determined independently of interferences and computed through an
integration over the distance covered by the maximum range and over
asymmetry variations). So, the difference between the two approaches
is less significant. The same interpretation can be used for
reliability evolution, presented in Fig. \ref{fig3}. Reliability
also undergoes the frame discards due to the reaching of maximum
frame retries or maximum CSMA backoffs. The rejected frames due to
full node queue represent also a possible interpretation with high
offered loads. The throughput evolution, presented in Fig.
\ref{fig4}, also undergoes the constraints of the PHY layer, and is
logically less significant since it follows the same evolution of
reliability (throughput is the product of reliability, offered load
and data frame size).

\subsection{Evolution of node performance with growing densities}
We use our model to compare node performances with three densities.
We propose a network with 5 nodes, another with 10 nodes and a third
with 50 nodes. We take into account the same outputs cited in the
previous section. The major observation is that the IEEE 802.15.4
networks do not support heavy traffic. The denser the network is,
the poorer are the performances are. We note an increasing delay for
denser networks, as observed in Fig. \ref{fig5}. As the number of
nodes increases, and with growing offered loads, collisions are more
frequent and so the retransmissions are more recurrent. The
switching phase to saturated network shows more significant
differences between the three network scenarios. Each node queue
begin to experience congestion problems; with more retransmission
requirements, the queues are busier and the delays are longer. At
saturation, the frame losses are widespread (collisions, link
constraints, frames discarded due to retry limits, etc.) for the
three scenarios, but the number of nodes still has an impact because
it has a negative influence on performances and channel availability
(more collisions, more retransmissions, channel congestion,\ldots).
The same reasoning explains a higher failure probability, as
presented in Fig. \ref{fig6} and a lower reliability as outlined in
Fig. \ref{fig7} for denser networks and with increasing offered
load. The evolution of throughput, shown in Fig. \ref{fig8}, also
matches with the interpretations cited above.

\section{Conclusion}
We have presented, in this paper, a model that mimics the IEEE
802.15.4 functionalities at the PHY and the MAC layers. We aim to
combine two relevant propositions. On the one hand, we model
constraints that affect link quality using the Zuniga and
Krishnamachari mathematical framework: distance, output power,
noise, asymmetry and errors related to encoding and modulation. The
PHY model bypasses the disk-shaped node range and expresses more
precisely the degree of link unreliability. The output of this model
represents an important outcome for estimating more faithfully the
probability of transmitting frame failure. On the other hand, we
enhance the Park et al. IEEE 802.15.4 MAC layer model to extract the
delay and the reliability. Our contribution seeks to improve the
Park et al. approach in determining inherent probabilities (frame
transmission, free channel in CCA1/CCA2, failure,...) and combining
it with the PHY model to better estimate wireless network
parameters. The methodology adopted relies on a Markov chain that
follows the flowchart described in Fig.\ref{Flowchart} and on an
M/M/1/K queue that we includes with the Park et al. approach. The
joint model is available at the SGIP NIST Smart Grid Collaboration
website \cite{NIST} for use.

The simulations show that more precise estimations are provided by
our model versus that by the Park et al. MAC model. The comparison
between the two considerations highlights a notable performance
deterioration using the combined model. This observation is quite
logical since this combination joins PHY constraints to collisions.
Thus, our contribution improves the Park et al. approach by
bypassing the assumption that failures are restricted to collisions.
The amelioration of the Park et al. approach is not limited to the
above description. We try also to enhance the estimation of inherent
probabilities by adjusting some expressions (as for $\alpha$,
$\beta$ and $P_{fail}$) and modifying the resolution method to
gather new parameters (such as $p_0$, the probability that a node
returns to the idle state, which is considered as an input in Park
et al. work).

Our contribution proposes to mimic the IEEE 802.15.4 PHY and MAC
layers mechanisms. Nonetheless, it is extensible for reproducing
more precise wireless networks standards related to IEEE 802.15.4.
It is also adjustable to other standards. Indeed, the considered PHY
layer model is quite relevant but assumes that interferences are
weak and/or stable. Moreover, the probability of an error at the PHY
layer is averaged (through integration over maximum range and
maximum asymmetry variation). Hence, as future work, we will seek to
extend our model in order to consider distance between nodes and
thereby topologies. Also, it will be challenging to plan a model
that deals with node mobility to appreciate its impact on
performances. From there, our model can be used as a tool to verify
metrics efficiency in mobile environments.

\bibliography{Biblio}
\bibliographystyle{ieeetr}

\end{document}